\documentstyle[12pt] {article}
\begin {document}
\parindent=15pt
\begin{center}
\vskip 1.5 truecm
{\bf MULTIPLE SOFT PION PRODUCTION WITHIN NONLINEAR CHIRAL SIGMA
MODEL}\\ \vspace{1cm} A.Shuvaev\\ Theory Department, St.Petersburg
Nuclear Physics Institute\\ 188350, Gatchina, St.Petersburg, Russia.\\
e-mail: {\tt shuvaev@thd.pnpi.spb.ru}
\end{center}
\begin{abstract}
Multiple soft pion production in the baryon scattering reactions is
considered in the framework of
 chiral nonlinear sigma model neglecting the pion mass. Treating
baryons in the eikonal approximation as classical sources, a set of
analytical solutions for the pion field is found. A tree $S$-matrix is
constructed on the basis of these solutions describing the
emission (or absorption) of any number of soft pions. Then the
contribution of soft virtual pions is taken into account in a closed
form. It is shown that the loop corrections strongly suppress the pion
radiation, and for the two limiting cases of nonrelativistic and
ultra-relativistic baryon scatterings there is no pion emission from
the "ends". Thus, the mechanism similar to the soft photon
bremsstrahlung in the quantum electrodynamics seems to be unable to
create a state with a large number of the soft pions.
\end{abstract}
\vspace{1cm}

\noindent
{\bf 1.} The problem of possible production of the
classical pion field in the high energy hadron-hadron or
ion-ion collisions has recently drawn much attention \cite{AnsRys}.
Very interesting example of the classical pion field
is a disoriented chiral condensate (DCC) introduced
by Bjorken and co-authors \cite{Bj}. Possible experimental
signatures of  DCC are of
special importance since one has to distinguish between
this phenomenon and the effects arising due to the
independent emission of a large number of soft pions.
It is well-known that these effects can be described
in terms of a coherent state resembling bremsstrahlung
photon in the electrodynamics \cite{W,B} (see also \cite{ABL}).
Neglecting the
pion interaction, such an approach leads
to the predictions for the charge distribution of pions
in the multiple production processes, which are really almost
identical to what could be expected from  DCC \cite{IA,IBNS}. That
is why it is of great interest to study how the pion
interaction modifies these results.

The multiple soft pion production in the hadron-hadon collisions
is treated here within the nonlinear sigma-model. This model
includes three isovector fields $\pi_i$, $i=1,2,3$, and an auxiliary
scalar field $\sigma$ obeying the constraint
$$
\sigma^2\,+\,\vec{\pi}^2\,=\,f_\pi^2, \qquad f_\pi = 93\, {\rm MeV}.
$$
The lagrangian is
\begin{equation}
\label{L}
L\,=\,L_\pi\,+\,L_f,
\end{equation}
where $L_\pi$ is a pure pionic part,
\begin{equation}
\label{Lpi}
L_\pi\,=\,\frac 12 \bigl[\,(\partial_\mu \sigma)^2\,+\,
(\partial_\mu \vec{\pi})^2\,\bigr],
\end{equation}
and $L_f$ describes the pion interaction with fermions,
$$
L_f\,=\,\overline{q}\, i\gamma^\mu \partial_\mu q\,-\,
g\overline{q}\bigl(\sigma + i\vec{\pi}\vec{\tau}\gamma_5\bigr)q.
$$
The vacuum of the model corresponds to the expectation values
$\langle\sigma\rangle = f_\pi$, $\langle\pi_i\rangle = 0$, which
result into the fermion mass $M = g f_\pi$. The vacuum is degenerate,
there is a set of vacuum configurations
$$
\langle\sigma\rangle = f_\pi\,\cos\,\theta, \quad
\langle\pi_i\rangle = f_\pi n_i\,\sin\,\theta
$$
for arbitrary parameter $\theta$ and unit vector $n_i$.

The model (\ref{L}) provides a simple but qualitatively satisfactory
description for soft pion-nucleon physics, $g_{\pi NN}$ and $M$ being
pion-nucleon interaction constant and  nucleon mass,
respectively.

The sigma model can be also  regarded as an effective chiral theory
describing the low energy limit of QCD. It correctly reflects the
important property of QCD, namely, spontaneous chiral symmetry breaking
due to the quark condensate $\langle\overline{q} q\rangle \not= 0$
and the goldstone nature of pions. In this case $M = M_q$ is the mass
of the constituent quark $q$.

The pion field represents the chiral phase of the quark
condensate, that is why the parametrization in the form
of  a unitary matrix
is quite natural in this approach. Constructing the matrix
$$
U\,=\,\frac 1{f_\pi}\,\bigl(\,\sigma\,+\,i\vec{\pi}\vec{\tau}\,
\bigr), \qquad U^+U\,=\,1,
$$
the lagrangian (\ref{Lpi}) takes the form
$$
L_\pi\,=\,\frac14 f_\pi^2\,Tr\bigl(\partial_\mu\,U^+\partial_\mu\,U
\bigr).
$$
It is obviously symmetric under the chiral transformation
$$
U\,\to\,S\,U\,S
$$
and the isotopic rotation
$$
U\,\to\,V^{-1}\,U\,V
$$
for any unitary matrices $S$ and $V$.

In this paper the mass $M$ is supposed to be much larger than the
pion momenta. This is the case of a soft pion bremsstrahlung when
pions are emitted from the ends of the diagram. Each external line
acquires, due to the pion emission, the factor
\begin{equation}
\label{fout}
\exp\left\{\frac i{2M}\,\int_0^\infty d\tau\,A(x_\tau)\right\}
\end{equation}
for outgoing particles and
\begin{equation}
\label{fin}
\exp\left\{\frac i{2M}\,\int_{-\infty}^0 d\tau\,A(x_\tau)\right\}
\end{equation}
for incoming ones. Here
$$
A(x)\,=\,g\bigl(\sigma(x)-f_\pi\bigr)\,\bar N (p)\,N(p)\,+\,
i\pi_i(x)\,\bar N (p)\,\tau_i\gamma_5\,N(p)\,=\,
2M\,g\bigl(\sigma(x)-f_\pi\bigr)
$$
and the integration goes along a worldline of an emitting particle
$x_\tau = \tau p/M $. These expressions are clearly equivalent to the
effect of a pointlike external source moving with a constant velocity
$\vec v = \vec p/M$, with a lagrangian
$$
L_s\,=\,g\bigl(\sigma(x)-f_\pi\bigr)\,\theta(\pm t)\,\delta_v^3
(\vec x-\vec v t),
$$
in which
$$
\delta_v^3(\vec x-\vec v t)\,=\,\delta\bigl(\frac{x_3 - v t}{\sqrt{1-v^2}}
\bigr)\,\delta^2(x_\perp).
$$
In matrix notations it is
\begin{equation}
\label{Lsm}
L_s\,=\,\frac 14 gf_\pi\,Tr\bigl(U\,+\,U^+-2\bigr)\,\delta_v^3
(\vec x-\vec v t).
\end{equation}
Detailed structure of colliding baryons is not important for
the soft emission which depends on the interaction constant $g$
only, which has a meaning of the baryon form factor. It can be expressed
through "scalar density" of the baryon state $|N\rangle$
$$
\rho(x)\,=\,\langle N|\,\overline{q}(x) q(x)\,|N\rangle
$$
as
\begin{equation}
\label{g}
g\,=\,\int d^3 x\,\rho(x).
\end{equation}
Here the pseudoscalar density $\rho_{\rm ps}=
\langle N|\,\overline{q}(x)\tau_i\gamma_5 q(x)\,|N\rangle$ is supposed to be
zero. From this point of view both interpretations of the sigma model
mentioned above give the same answer. However, this is not true for the
central part of the hard process
occuring at small distances for $t=0$.

\smallskip
\noindent
{\bf 2.} To begin with, consider the classical pion field produced by
a single point-like source moving with constant velocity $\vec v$ along the
$x_3$ direction. Upon varying the lagrangian with respect to the pion
matrix $U$, the equation of motion yields
\begin{equation}
\label{dUdUdlt}
\partial_\mu\,\bigl[U^+\partial_\mu U\bigr]\,=\,\frac g{2 f_\pi}\,
\delta_v^3(\vec x-\vec v t)\,\bigl[U^+- U\bigr].
\end{equation}
In what follows the matrix $U(x)$ is sought among exact
solutions found in refs. \cite{Ans,AnsBan}:
\begin{equation}
\label{ans}
U(x)\,=\,V^{-1}e^{i\tau_3 f(x)}\,V,
\end{equation}
where $V$ is an arbitrary but constant unitary matrix. The function
$f(x)$ obeys the equation
$$
\partial^2 f(t,\vec x)\,=\,\frac g{f_\pi}\,\delta_v^3(\vec x-\vec v t)
\,\sin\, f(t,\vec v t).
$$
The solution decreasing at the infinity has the
 form of  "moving Coulomb
potential":
\begin{eqnarray}
\label{ftx}
f(t,\vec x)\,&=&\,-\frac g{4\pi f_\pi}\,\frac 1{\bigl|x_v\bigr|}\,
\sin\, f(\tau_v,\vec v \tau_v), \\
\bigl|x_v\bigr|\,&=&\,\left[\left(\frac{x_3-vt}{\sqrt{1-v^2}} \right)^2
\,+\,x_\perp^2\right]^{\frac 12}, \nonumber\\
\label{tauv}
\tau_v\,&=&\,\frac 1{1-v^2}\left[t\,-\,vx_3\,-\,\sqrt{1-v^2}\,
\bigl|x_v\bigr|\right]
\end{eqnarray}
which is defined through unknown function $f(\tau_v,\vec v \tau_v)$.
To get the closed equation for it one has to take the l.h.s. of (\ref{ftx})
at the source worldline $(t,\vec v t)$. However $f(t,\vec v t)$ is
divergent since $|x_v|$ turns into zero. As the divergency is a clear
consequence of the point-like structure of the source, it is straightforward
to cure it by spreading the delta-function in  eq. (\ref{dUdUdlt}) over a
small but finite space volume:
\begin{equation}
\label{dUdUrho}
\partial_\mu\,\bigl[U^+\partial_\mu U\bigr]\,=\,\frac g{2 f_\pi}\,
\rho_v(\vec x-\vec v t)\,\bigl[U^+- U\bigr],
\end{equation}
where the density
$$
\rho_v(\vec x-\vec v t)\,=\,\rho\bigl(\frac{x_3-vt}{\sqrt{1-v^2}},
x_\perp \bigr)
$$
is normalized according to (\ref{g}).

The function $f(t,\vec x)$ is generally related to the density
through the Green function,
\begin{equation}
\label{grint}
f(t,\vec x)\,=\,-\frac 1{4\pi f_\pi}\,\int\,d\tau d^3y\,\theta(t-\tau)
\delta\bigl[(t-\tau)^2-(\vec x-\vec y)^2\bigr]\,\rho_v(\vec y-\vec v\tau)\,
\sin\,f(\tau,\vec y).
\end{equation}
The retarded Green function chosen here ensures $f(t,\vec x)\to 0$
for $t\to\pm\infty$, that is the absence of the emission or absorbtion
of real particles. If the distance $x$ is larger than the size of the
source, one can replace $\vec y$ in the delta-function by $\vec v\tau$,
and the integral (\ref{grint}) results into eq. (\ref{ftx}).

For the points on the trajectory the integral is
$$
f(t,\vec v t)\,=\,\frac 1{4\pi f_\pi}\,\int\,d\tau d^3y\,\theta(\tau)\,
\delta\bigl[\tau^2-(\vec v\tau-\vec y)^2\bigr]\,\rho_v(\vec y)\,
\sin\,f\bigl(t-\tau, \vec y+\vec v (t-\tau)\bigr),
$$
or, after integrating over $\tau$,
$$
f(t,\vec v t)\,=\,\frac 1{4\pi f_\pi}\int\,d^3y\,\frac 1{2\sqrt(1-v^2)}
\frac 1{|y_v|}\,\rho_v(\vec y)
\sin\,f\bigl(t-\tau, \vec y+\vec v (t-\tau)\bigr).
$$
Keeping here the most singular terms at $y\to 0$, the equation for
$f(t,\vec v t)$ takes the form
\begin{equation}
\label{fsinf}
f(t,\vec v t)\,=\,-\frac 1a \sin\,f(t,\vec v t),
\end{equation}
where the parameter $a$,
$$
\frac 1a\,=\,\frac 1{8\pi f_\pi}\,\int\,d^3y\,\frac{\rho(\vec y)}{|y|},
$$
is proportional to the radius of the source.

Since $f(t,\vec v t)$ is supposed to be a smooth function, the solutions
of eq. (\ref{fsinf}) are time-independent constants $f(t,\vec v t)=f_n$.
When $a\to 0$,
$$
\sin\,f_n\,\simeq \pi a n,\quad n\,=\,0,\pm 1,\pm 2,\ldots,\quad
|n|\le \frac 1{\pi a}.
$$
Finally, there is a set of solutions which at large distances are
$$
f(t,\vec x)\,=\,n\,\frac r{|x_v|},\quad n\,=\,0,\pm 1,\pm 2,\ldots,\quad
|n|\le \frac g{4\pi f_\pi r},
$$
where $r= g/4f_\pi a$ is the effective radius of the source.

Notice that in contrast to the pure pionic case the solution (\ref{ans})
contains only one arbitrary unitary matrix. This loss of generality has
an obvious explanation. While the interaction (\ref{Lsm}) allows for the
independent isotopic rotation of the pion field irrespective of the
source, the chiral transformation has to be accompanied by an
appropriate transformation of the source densities. The "chiral phase"
of the solution (\ref{ans}) is fixed by the conditions $\langle
\overline q q\rangle \not =0, \langle \overline q \gamma^5 \vec\tau
q\rangle=0$ imposed on the source.  After  chiral transformation it
becomes a solution of the pion field equation with chirally
transformed density.

\smallskip
\noindent
{\bf 3.} Using exact solutions described above, consider
the $S$-matrix for the soft pion production.
Generally, the $S$-matrix is a functional depending on the pion
fields:
\begin{equation}
\label{Sg}
S[\pi^0]\,=\,\sum_{n\ge 0}\int\,dx_1\cdots dx_n\,S^{(n)}
_{i_1,\ldots,i_n}
(x_1,\ldots,x_n)\,\pi_{i_1}^0(x_1)\cdots \pi_{i_n}^0(x_n).
\end{equation}
The $\pi^0$ fields are free,
$$
\partial^2\,\pi_i^0(x)\,=\,0,
$$
and expressed through the pion creation and annihilation operators
\begin{equation}
\label{pi0}
\pi_i^0(x)\,=\,\frac 1{(2\pi)^{3/2}}\int\,\frac{d^3 k}{2 k_0}\,
\bigl[e^{ikx}\,a_i^+(k)\,+\,e^{-ikx}\,a_i(k)\bigr],
\end{equation}
these operators being normally ordered in eq. (\ref{Sg}).

Cross section for $N$ emitted  pions which carry the momenta
$k_1,\ldots, k_N$ is given by the matrix element
$$
d\sigma_{i_1,\ldots,i_N}^{(N)}(k_1,\ldots,k_N)\,=\,\bigl|\langle\, 0\,|
\,a_{i_1}(k_1)\cdots a_{i_N}(k_N)\,S[\pi_0]\,|\,0\,\rangle\bigr|^2 d\tau_N.
$$
Here $|0\rangle$ is the vacuum state and $d\tau_N$ is the phase volume.
The distribution of pions can be evaluated as
\begin{equation}
\label{mult}
d\overline N_{i_1,\ldots,i_N}(k_1,\ldots,k_N)\,=\,
\frac{\sum_{N=1}^\infty N d\sigma_{i_1,\ldots,i_N}^{(N)}(k_1,\ldots,k_N)}
{\sigma_{tot}},
\end{equation}
where the total cross section is
$$
\sigma_{tot}\,=\,\sum_N \sum_{i_1,\ldots,i_N} \int d\sigma_{i_1,\ldots,i_N}^
{(N)}(k_1,\ldots,k_N).
$$

According to general rules for the $S$-matrix construction in the
tree
approximation, one has to find the solution $\pi_i(x)$ of the classical
equation, which is a free field for $t\to\pm\infty$.
The positive-frequency part of the
asymptotics for $t=+\infty$ and the negative one
for $t=-\infty$ have to be equal, respectively, to the positive and
negative parts of the field $\pi^0$ (\ref{pi0}). Then the tree
$S$-matrix is evaluated through the classical action calculated
with this solution:
$$
S[\pi^0]\,=\,{\cal N}\,\exp\,\left\{i\int\,d^4x\,L(\pi) \right\}.
$$
${\cal N}$ is a normalization constant. Here
the action  should be regularized by subtracting the asymptotic
$\pi_i^0$ from the field $\pi_i$ in the quadratic terms:
$$
\frac 12\,\partial_\mu\pi_i\partial_\mu\pi_i\,\to\,
\frac 12\,\partial_\mu\bigl(\pi_i-\pi_i^0\bigr)
\partial_\mu\bigl(\pi_i-\pi_i^0\bigr).
$$
This way of proceeding follows from the saddle-point approximation for
the functional $S$-matrix integral \cite{Fad}.

For the soft pion momenta, the terms containing the field derivatives
can be omitted, and the tree $S$-matrix is expected to be of the local
form,
\begin{equation}
\label{lS}
S[\pi^0]\,=\,\exp\left\{\sum_{n\ge 0}\int\,dx\,
S^{(n)}_{i_1,\ldots,i_n}(x)\,\pi_{i_1}^0(x)\cdots \pi_{i_n}^0(x)
\right\},
\end{equation}
or of the sum of exponents of this type, if there are several solutions
of the classical field equations. The isotopic invariance implies
the independence of the $S$-matrix of the direction of
$\vec\pi_0$ in the isotopic space, $S=S(|\vec\pi_0(x)|)$.
This property enables us to recover the $S$-matrix at the tree level
through classical solutions (\ref{ans}), with the
constant isotopic orientation.

\smallskip
\noindent
{\bf 4.} For definitness the annihilation
of two nucleons into pions is taken as an example. The colliding
baryons are described in the centrer-of-mass system as
the two sources moving
towards each other from the space-time infinity. They give rise to the
classical pion field obeying the equation
$$
\partial_\mu\,\bigl[U^+\partial_\mu U\bigr]\,=\,\frac g{2 f_\pi}\,
\theta(-t)\,\bigl[\delta_v^3(\vec x-\vec v t)-
\delta_v^3(\vec x+\vec v t)\bigr]\,\bigl(U^+- U\bigr),
$$
or, in terms of the function $f$ in the parametrization (\ref{ans})
\begin{equation}
\label{d2f2}
\partial^2f\,=\,\frac g{ f_\pi}\,\theta(-t)\,\bigl[\delta_v^3(\vec x-
\vec v t)-\delta_v^3(\vec x+\vec v t)\bigr]\,\sin\,f.
\end{equation}
Similarly to a single source, the general solution of eq. (\ref{d2f2})
can be written as a sum of two "moving Coulomb potentials":
\begin{eqnarray}
f(t,x)&=&\varphi_0(t,x)\,-\,\frac g{4\pi f_\pi}\,\theta(-\tau_v)\,
\frac 1{\bigl|x_v\bigr|}\,\sin\, f(\tau_v,\vec v \tau_v)\,+\nonumber\\
&+&\frac g{4\pi f_\pi}\,\theta(-\tau_{-v})\,
\frac 1{\bigl|x_{-v}\bigr|}\,\sin\, f(\tau_{-v},\vec v
\tau_{-v}),\nonumber
\end{eqnarray}
with so far unknown functions $f(\tau_{\pm v},\vec v \tau_{\pm v})$
of the proper time (\ref{tauv})
and the function $\varphi_0$ satisfying the free equation,
$\partial^2\varphi_0=0$. For $t\to\pm\infty$ (and fixed $\vec x$)
$f(t,x)\to\varphi_0(t,x)$, so it is the function $\varphi_0$ which defines
the asymptotic pion field, $\pi_0(t,x)=f_\pi\varphi_0(t,x)$.

To obtain the closed equations for $f$, one has to take into
account the finite sizes of the sources, that is to replace the
delta-functions in eq. (\ref{d2f2}) by more smooth densities
$\rho_v(x\pm vt)$ localized in a small volume. This leads to the following
equations for the function $f$ at the wordlines of the colliding particles
\begin{eqnarray}
&&f(t,\vec v t)\,-\,\varphi_0(t,\vec v t)\,=\nonumber\\
&=&-\frac 1a\,\left[\sin\,f(t,\vec v t)\,-\,\frac g{4\pi f_\pi}\,
\sqrt{1-v^2}\frac a{2v|t|}\,\sin\,f\bigl(\frac{1-v}{1+v}t,-\vec v
\frac{1-v}{1+v}t\bigr)\right]\nonumber\\
\label{ftvt}
&&\\
&&f(t,-\vec v t)\,-\,\varphi_0(t,-\vec v t)\,=\nonumber\\
&=&-\frac 1a\,\left[\sin\,f(t,-\vec v t)\,-\,\frac g{4\pi f_\pi}\,
\sqrt{1-v^2}\frac a{2v|t|}\,\sin\,f\bigl(\frac{1-v}{1+v}t,\vec v
\frac{1-v}{1+v}t\bigr)\right].\nonumber
\end{eqnarray}
Second terms in the r.h.s. of (\ref{ftvt}) are negligible
at large distances, and
they become of the same order as the first ones only when the distance
between the sources is comparable with their size, $v|t|\sim r$.
However the region, where the colliding particles overlap,
corresponds to the central part of the process related to the
hard stage of the reaction. Here the dynamics is governed by the
detailed structure of baryons, which is not incorporated in the
model of soft pion emission by classical sources the present approach
is based on. This is the reason why the second terms are to be dropped
in eq. (\ref{ftvt}), and the equation determining the soft pion
bremsstrahlung amplitude is
$$
f(t,\pm\vec v t)\,-\,\varphi_0(t,\pm\vec v t)\,=\,\mp\,\frac 1a\,
\sin\,f(t,\pm\vec v t).
$$

For small $a$, one gets the two sets of solutions similar to those
obtained for a single source:
\begin{eqnarray}
\sin\,f_n(t,vt)&=&-a\bigl(\pi n\,+\,\varphi_0(t,vt)\bigr),\nonumber \\
&& n\,=\,0,\pm 1,\pm 2,\ldots,\quad |\pi n-\varphi_0|\le \frac 1a,
\nonumber \\
\label{fnfm}
&& \\
\sin\,f_m(t,-vt)&=&a\bigl(\pi m\,+\,\varphi_0(t,-vt)\bigr),\nonumber \\
&& m\,=\,0,\pm 1,\pm 2,\ldots,\quad |\pi m-\varphi_0|\le \frac 1a.
\nonumber
\end{eqnarray}

Now one can calculate the tree $S$-matrix substituting this solution by
the classical action. As was mentioned above, one has to regularize
the piece of the action which gives rise to the free pion propagator.
Separating this term explicitly, the modified action density can be
written as
\begin{eqnarray}
\tilde L &=& \frac 12\,\partial_\mu\bigl(\pi_i-\pi_i^0\bigr)
\partial_\mu\bigl(\pi_i-\pi_i^0\bigr)\,+\,
\frac14 f_\pi^2\,Tr\bigl(\partial_\mu\,U^+\partial_\mu\,U \bigr)\,-\,
\frac 12\,\partial_\mu\pi_i\partial_\mu\pi_i \,+\nonumber\\
&+& \frac 14 gf_\pi\,\bigl[\delta_v^3(\vec x-\vec v t)-
\delta_v^3(\vec x+\vec v t)\bigr]\,Tr\bigl(U\,+\,U^+-2\bigr),\nonumber
\end{eqnarray}
$$
U\,=\,e^{i \vec\pi \vec\tau}.
$$
Then the tree $S$-matrix is given by the sum of the terms calculated
with this action for each solution of the pion field equation (\ref{fnfm})
\begin{eqnarray}
S\,=\,&{\cal N}&\,\sum_{n,m}\exp\,igf_\pi\left\{\int_{-\infty}^0\,dt\,
\bigl[\frac 1{2a}
\,\sin^2f_n + \cos\,f_n \bigr]\,\right.-\nonumber\\
&-&\left. \int_{-\infty}^0\,dt\,\bigl[\frac 1{2a}
\,\sin^2f_m + \cos\,f_m \bigr]\,\right\}.\nonumber
\end{eqnarray}

Using the explicit expression (\ref{fnfm}),
consider the contribution to this sum coming from the first particle:
\begin{eqnarray}
S_1&=&{\cal N}\,\sum_n\exp\biggl\{igf_\pi\,\int_{-\infty}^0 dt
\biggl[\frac 12\,a\bigl(\pi n-\varphi_0 \bigr)^2\,-\nonumber \\
&-&(-1)^n\,\sqrt{1-a^2\bigl(\pi n-\varphi_0 \bigr)^2}\,\biggr]\biggr\}.
\nonumber
\end{eqnarray}
For small $a$ the sum can be converted into two integrals over the
variable $z=\pi a n$, the odd and even $n$ being taken separately.
Keeping only the  terms nonvanishing at $a\to 0$, these integrals take
the form
\begin{eqnarray}
S_1&=&{\cal N}\,\int_{-1}^1
dz\,\exp\left\{igf_\pi\,\int_{-\infty}^0 dt\, \left[\frac
1{2a}\,z^2\,-\,z\,\varphi_0\,+\,\sqrt{1-z^2}\right]\right\}\,+
\nonumber\\
&+&{\cal N}\,\int_{-1}^1 dz\,\exp\left\{igf_\pi\,\int_{-\infty}^0 dt\,
\left[\frac 1{2a}\,z^2\,-\,z\,\varphi_0\,-\,\sqrt{1-z^2}\right]\right\}.
\nonumber
\end{eqnarray}
The first and the last terms in the exponents are divergent due to the
time integrals. Regularizing them by the long-distance cutoff $T$, the
first integral
$$
{\cal N}\,\int_{-1}^1 dz\,\exp\,igf_\pi\,T\left[\frac 1{2a}\,z^2\,+\,
\sqrt{1-z^2}
\right]\,\exp\left\{-igf_\pi\,z\,\int_{-\infty}^0 dt\,\varphi_0\right\}
$$
can be calculated for $T\to\infty$ through the stationary point which
is at $\sqrt{1-z^2}=a$. Since there is no stationary
point on the real axis of the complex plane $Z$ for the second
integral, it vanishes for large $T$ \footnote{There is another
solution $z=0$ for both these integrals. However, it results in the
constant independent of the field $\varphi_0$.}. Collecting all the
factors, which do not contain the field $\varphi_0$, into the overall
constant ${\cal N}$, one gets a simple result:
\begin{equation}
\label{S1phi}
S_1\,=\,{\cal N}\,\sum_{\kappa=\pm}\,\exp\left\{igf_\pi\,\kappa\,
\int_{-\infty}^0 dt\,\varphi_0(t,vt)\right\}.
\end{equation}

Recall now that due to the isotopic invariance the $S$-matrix
actually
depends only upon the absolute value of the isotopic vector $\vec\pi_0(x)$
and that eq. (\ref{S1phi}) is obtained for a particular direction of this
vector. So  one can rewrite it as
\begin{equation}
\label{Sin}
S_{in}\,=\,{\cal N}\,\sum_{\kappa=\pm}\,\exp\left\{ig\,\kappa\,
\int_{-\infty}^0 dt\,\sqrt{\vec\pi_0^2}\right\}.
\end{equation}

The generalization for any number of colliding particles is straightforward
-- each incoming line has to be supplemented with the factor
(\ref{Sin}), while the outgoing one acquires the factor
\begin{equation} \label{Sout} S_{out}\,=\,{\cal
N}\,\sum_{\kappa=\pm}\,\exp\left\{ig\,\kappa\,\int_0^\infty dt\,
\sqrt{\vec\pi_0^2}\right\},
\end{equation}
where the integrals go along the particle trajectories.

Notice that the square roots here are due to the
nonlinear character of pion field. In the absence of the pion
self-interaction the expressions (\ref{fout}), (\ref{fin}) would produce
the factors
$$
S\,=\,{\cal N}\,\exp\biggl\{\pm ig\,\int\,dt\,\sqrt{f_\pi^2
- \vec\pi_0^2}\,\biggr\}
$$
(due to the property $\langle \tau_i \gamma_5\rangle =0$ only
even powers of the pion field contribute). Thus, at least for a weak field,
nonlinear effects enhance the amplitude at the tree level.

\smallskip
\noindent
{\bf 5.} The functions (\ref{Sin}), (\ref{Sout}) give the $S$-matrix
in the tree
approximation. A natural question is to what extent the loop corrections
change this result. As far as the effective low energy theory is
treated, the loop momenta have to be cut off from above by a typical
scale of the order of nucleon mass. To find the contribution from soft
virtual pions one can use an approach similar to the equivalent photon
method of the electrodynamics. Assuming the vertices for the soft
virtual and real pion emission to be the same, one can yield the soft
loops connecting the incoming and outgoing lines of the tree $S$-matrix
by the pion propogator $1/k^2$. Then the $S$-matrix accounting for all
the loops can be written as
the functional averaging of the tree $S$-matrix
over  virtual pion field $\pi_i(x)$
\begin{eqnarray}
\label{Sloop}
S[\pi^0]\,&=&\,{\cal N}\,\prod_{\kappa_i}\,\int\,D\pi_i(x)\,\exp i \biggl\{
\int\,d^4x\, \bigl[\,\frac 12 \,(\partial_\mu \vec\pi)^2\,+ \\
&+&\,g\kappa_i\,
\theta(\pm t)\,\delta_v^3(x-v_it)\,\sqrt{(\vec\pi_0 + \vec\pi)^2}\,\bigr]
\biggr\},
\nonumber
\end{eqnarray}
where the product is taken according to eqs. (\ref{Sin}), (\ref{Sout})
over the external lines of the basic diagram.
Actually, eq. (\ref{Sloop}) is a special case of the general recipe
for the $S$-matrix expressed through a functional integral
used above. The integral (\ref{Sloop}) effectively re-sums the quantum
fluctuations around the classical solution.

With the tree $S$-matrix linear in the pion field and without a square
root, $S\sim \exp \int J_i(x)\pi_i^0(x) dx$ ($J_i$'s are  external
sources), this integral would be of the form
\begin{eqnarray}
S[\pi^0]\,&=&\,{\cal N}\,\int D\pi_i(x)\,\exp i \biggl\{
\int\,d^4x\, \bigl[\,\frac 12\, (\partial_\mu \pi_i)^2\,+ \,J_i(\pi_i +
\pi_i^0)\,\bigr]\biggr\}\,= \nonumber\\
&=&\,{\cal N^\prime}\,\exp\,\biggl\{\int\,d^4x\,J_i \pi_i^0\biggr\},\nonumber
\end{eqnarray}
and the loop corrections would be completely absorbed by the normalization
${\cal N^\prime}$. This is the case of the quantum
electrodynamics where the soft virtual photons change the cross section but
do not affect the normalized distribution of the type of (\ref{mult}).

To get rid of the square root in eq. (\ref{Sloop}) one can represent
it through the functional integral over auxiliary fields $\vec n_i(t)$
and $\alpha(t)$ $$ e^{i
g\kappa\,\int_{-\infty}^0\sqrt{\vec\pi^2}dt}\,=\, {\cal N}\,\int\,Dn
D\alpha\,\exp i \biggl\{\int_{-\infty}^0 dt \bigl[g \vec n_i \vec
\pi_i\,-\,\frac 12\alpha\kappa \bigl(\vec n^2-1\bigr)\bigr]\biggr\}, $$
the variable $\alpha(t)$ taking
 positive values only, $\alpha(t)>0$.
This formula is easily derived by discretizing the time interval in the
l.h.s. and making the Fourier transform for  each moment of time. Here
the functional measure $Dn D\alpha$ denotes  the product of
discrete integrals:
\begin{equation} \label{measure} Dn
D\alpha\,=\,\prod_t\int_{-\infty}^\infty d^3n_t\int_0^\infty d\alpha_t
e^{-\alpha_t\delta},
\end{equation}
where $\delta\to+0$ ensures the convergency of the $\alpha_t-$integrals
in the upper limit.

Now the integrals over  virtual pions turn out to be of a gaussian type
and can be calculated.
Returning again to the case of the two colliding baryons,
the result is
\begin{equation}
\label{fint}
S[\pi_0]\,=\,{\cal N}\,\int\prod_{i=1}^2 Dn_i D\alpha_i\,e^{-i
\int_{-\infty}^0 dt_i\bigl[ \frac 12 \alpha_i\kappa_i(\vec n_i^2-1) -
g\vec n_i\vec\pi_i^0 \bigr]}\,e^\Phi,
\end{equation}
where $\Phi$ can be symbolically written as
$$
\Phi\,=\,\frac 12 g^2\,\bigl[\vec n_1(t)\delta_v^3(x-vt)+
\vec n_2(t)\delta_v^3(x+vt)\bigr]\,\frac 1{\partial^2}\,
\bigl[\vec n_1(t)\delta_v^3(x-vt)+\vec n_2(t)\delta_v^3(x+vt)\bigr]
$$
and $1/\partial^2$ stands for the pion propagator. Substituting its
expression in the coordinate space as follows
\begin{equation}
\label{prop}
\langle x|\,\frac 1{\partial^2}\,|y\rangle\,=\,-\,\frac i{4\pi^2}\,
\frac 1{(x-y)^2 - i\delta},
\end{equation}
one arrives to the explicit form of  $\Phi$:
\begin{eqnarray}
\Phi\,&=&\,\frac{g^2}{8\pi^2}\,\int_{-\infty}^0 dt_1 dt_2\,
\biggl\{\sum_{i=1}^2 \vec n_i(t_1)\,\frac 1{(1-v^2)(t_1-t_2)^2 -i\delta}\,
\vec n_i(t_2)\,+ \nonumber\\
\label{Phi}
&+&\,2\,\vec n_1(t_1)\,\frac 1{(t_1-t_2)^2 -v^2(t_1+t_2)^2-i\delta}\,
\vec n_2(t_2)\biggr\}.
\end{eqnarray}

Starting from this point, the two limits will be taken separately: the
nonrelativistic limit, $v\approx 0$, and the ultra-relativistic one,
$v\approx 1$.

For $v\approx 0$ the function $\Phi$ takes the form
$$
\Phi_{\rm NR}\,=\,\frac{g^2}{8\pi^2}\,\int_{-\infty}^0 dt_1 dt_2\,
\sum_{i,k=1}^2\frac{\vec n_i(t_1)\,\vec n_k(t_2)}
{(t_1-t_2)^2 -i\delta}.
$$
The integrals in this expression are singular for $t_1=t_2$. They can be
regularized by the parameter $r$ which has a meaning of the ultraviolet
cutoff, $(t_1-t_2)^2\to (t_1-t_2)^2+r^2$. In terms of the variable
$\tau= (t_1-t_2)/2r$, the regularized integral is $$ \Phi_{\rm
NR}\,=\,\frac{g^2}{4\pi^2}\,\frac 1r\,\int_{-\infty}^0 dt
\int_{\tau/r}^0 d\tau\sum_{i,k=1}^2 \frac{\vec n_i(t)\,\vec n_k(t-r\tau)}
{\tau^2 +1},
$$
or, for small $r$,
\begin{equation}
\label{opK}
\Phi_{\rm NR}\,=\,\frac{g^2}{8\pi r}\,\int_{-\infty}^0 dt \sum_{i,k=1}^2
\vec n_i(t)\,\vec n_k(t)\,+\,\vec n\,K\,\vec n
\end{equation}
where all the terms, but the most singular ones, are collected in the
operator $K$
in such a way that $r\,K\to 0$ for $r\to 0$. The functional integral
over the auxiliary fields $\vec n_i(t)$ is of the gaussian type, with
the quadratic exponential part
$$ \frac 12 \int dt_1 dt_2 \sum_{i,j}
\vec n_i(t_1)\,F_{ij}(t_1,t_2)\,\vec n_j(t_2)
$$
given by the $2\times 2$ block matrix
\begin{equation}
\label{block}
F(t_1,t_2)\,=
\end{equation}
$$
=\,\left(\begin{array}{cc} \bigl( \frac ar - i\kappa_1\alpha_1
\bigr)\delta_{t_1,t_2}+K(t_1,t_2)&\frac ar \delta_{t_1,t_2}+K(t_1,t_2)\\
\frac ar \delta_{t_1,t_2}+K(t_1,t_2)&\bigl(\frac ar - i\kappa_2\alpha_2
\bigr)\delta_{t_1,t_2}+K(t_1,t_2) \end{array}\right),
$$
$$
a\,=\,\frac{g^2}{8\pi},\qquad \delta_{t_1,t_2}\,\equiv\,\delta(t_1-t_2),
$$
in terms of which the integral yields
\begin{equation}
\label{gauss}
Det\, F^{-\frac 32}\,\exp\,\int dt_1 dt_2\, \frac 12 g^2\sum_{i,j}
\vec \pi_i^0(t_1)\,F_{ij}^{-1}(t_1,t_2)\,\vec \pi_j^0(t_2).
\end{equation}
The inverse matrix $F^{-1}$ and its determinant are expressed through
the operator $K$ (\ref{opK}) as
$$
\bigl(F^{-1}\bigr)_{ik}\,=\,i\delta_{ik}\kappa_k\alpha_k^{-1}\,+\,
\kappa_i\kappa_k\alpha_i^{-1}\bigl( \frac ar + K\bigr)\bigr[I+i\sum_{j=1}^2
\kappa_j\alpha_j^{-1}\bigl( \frac ar + K\bigr)\bigr]^{-1}\alpha_k^{-1},
$$
$$
Det\, F\,=\,\prod_{j=1}^2 Det\,\alpha_j \cdot Det \bigr[I+i\sum_{j=1}^2
\kappa_j\alpha_j^{-1}\bigl( \frac ar + K\bigr)\bigr].
$$
Here $\alpha_j=\alpha_j(t)$ is understood as a multiplication operator.
As is seen from
the first line, the matrix $F^{-1}$ has a finite limit for $r\to 0$:
\begin{equation}
\label{Fr}
F^{-1}\begin{array}{c} \\ \to \\ ^{r\to 0} \end{array}  \frac 1 {\kappa_1
\alpha_1 + \kappa_2\alpha_2}\,\left(\begin{array}{rr} I& -I\\
-I&I\end{array}\right).
\end{equation}
In this limit
$$
Det\, F\,=\,c\,Det\,\bigl|\kappa_1\alpha_1 + \kappa_2\alpha_2 \bigr|,
$$
where all $\alpha-$independent  factors are included in the constant
$c$.

Thus, the $S$-matrix turns out to be finite for $r\to 0$:
\begin{eqnarray}
\label{Sr0}
S[\pi_0]\,&=&\,{\cal N}\sum_\kappa\int\prod_i D\alpha_i\,Det
\bigl|\kappa_1\alpha_1 + \kappa_2\alpha_2 \bigr|^{-\frac 32}\times\\
&\times& \exp\,i\biggl\{\frac 12 \int_{-\infty}^0 dt \biggl[\kappa_1\alpha_1
+ \kappa_2\alpha_2 + g^2\frac{(\vec\pi^0_1-\vec\pi^0_2)^2}
{\kappa_1\alpha_1 + \kappa_2\alpha_2}\biggr]\biggl\},\nonumber
\end{eqnarray}
$$
\pi^0_1=\pi_0(t,vt), \qquad \pi^0_2=\pi_0(t,-vt).
$$
Crucial point here is that the matrix $F(t_1,t_2)$ becomes
degenerate for a vanishing regulator. Really the result (\ref{Sr0}) does
not depend on the particular choice of the cutoff. Instead of
$r$ one can take, for instance, a finite value for the parameter $\delta$
in the propagator (\ref{prop}).

The integral (\ref{Sr0}) allows for a straightforward calculation
being a time
product of integrals of the same type,  without derivatives in the
exponent.  Its discretized version reads
\begin{eqnarray}
S[\pi_0]\,&=&\,{\cal N}\,\prod_t\,\int_0^\infty d\alpha_{1t} d\alpha_{2t}\,
e^{-\delta(\alpha_{1t}+\alpha_{2t})}\,|\kappa_1\alpha_{1t}+
\kappa_2\alpha_{2t})|^{-\frac 32}\times\nonumber\\
&\times&\,\biggl\{1\,+\,\frac 12 i\bigl[
\kappa_1\alpha_{1t}+\kappa_2\alpha_{2t}\,
+\,g^2\frac{(\pi_{1t}^0-\pi_{2t}^0)^2}{\kappa_1\alpha_{1t}+
\kappa_2\alpha_{2t}}\bigr]\,\Delta t \biggr\}.\nonumber
\end{eqnarray}
Or, after assigning a finite value to the parameter $\delta$ in the
measure (\ref{measure}),
one gets
$$ S[\pi_0]\,=\,{\cal N}\,\prod_t\,\sqrt\pi
\delta^{-\frac 12} \bigl\{1\,+\,i\kappa \bigl[\frac 14
\delta^{-1}\,-\,g^2 \delta (\pi_{1t}^0-\pi_{2t}^0)^2\bigr]\Delta
t\bigr\} $$ for the same signs, $\kappa_1=\kappa_2\equiv
\kappa$, and,
when the signs are different, $\kappa_1=-\kappa_2$, $$
S[\pi_0]\,=\,{\cal N}\,\prod_i\,2\sqrt{2\pi} \delta^{-\frac 12}.
$$
The other terms in the last line vanish due to the antisymmetry of the
integrand (here the principal value is taken for
the term $1/(\alpha_1-\alpha_2)$).

In the continuous limit the first line turns into the function
$$
S[\pi_0]\,=\,{\cal N}^\prime\,\exp\biggl\{i\kappa\,\int_{-\infty}^0 dt\,
\bigl[\frac 14 \delta^{-1}\,-\,g^2 \delta (\pi_1^0-\pi_2^0)^2\bigr]\biggr\}.
$$
Only the first term survives when $\delta \to 0$. However, this term
though strongly divergent is included into the constant ${\cal
N}^\prime$.  The second one is of the order
of $v^2$ and should be also dropped
by this reason, since the terms $v^2$ are neglected in eq.
(\ref{Phi}).  Thus, all the pieces of the $S$-matrix result into the
constants. Since the pion multiplicity  (\ref{mult}) does
not depend on the normalization, one can conclude that there is no soft
pion emission from the "ends" in the nonrelativistic limit. It is
completely suppressed by the loop corrections.

In the ultra-relativistic limit, $v_1\sim 1$, $v_2\sim -1$ the function
$\Phi$ (\ref{Phi}) takes the form
\begin{eqnarray}
\Phi_{\rm UR}\,&=&\,\frac{g^2}{8\pi^2}\,\int_{-\infty}^0 dt_1 dt_2\,
\biggl\{\sum_{i=1}^2 \vec n_i(t_1)\,\frac 1{(1-v_i^2)(t_1-t_2)^2 -i\delta}\,
\vec n_i(t_2)\,+\nonumber \\
&+&\,\vec n_1(t_1)\,\frac 1{1-v_1v_2}\,\frac 1{t_1 t_2-i\delta}\,
\vec n_2(t_2)
\biggr\}.\nonumber
\end{eqnarray}
Rewriting again the exponential part of
the integral (\ref{fint}) as
$$
\frac 12 \int dt_1 dt_2 \sum_{i,k}
\vec n_i(t_1)\, F_{v,ij}(t_1,t_2)\, \vec n_j(t_2),
$$
where
\begin{equation}
\label{Fv}
F_v(t_1,t_2)\,=
\end{equation}
$$
=\,\left(\begin{array}{cc} \bigl( \frac a{\beta_1 r} +  i\kappa_1\alpha_1
\bigr)\delta_{t_1,t_2}+\frac 1{\beta_1} K(t_1,t_2)&
\frac a{2\pi}\frac 1{1-v_1v_2}\frac 1{t_1 t_2-i\delta}\\
\frac a{2\pi}\frac 1{1-v_1v_2} \frac 1{t_1 t_2-i\delta}&
\bigl( \frac a{\beta_2 r} +
i\kappa_2\alpha_2\bigr)\delta_{t_1,t_2}+\frac 1{\beta_2} K(t_1,t_2)
\end{array}\right),
$$
$$
\beta_i\,=\,1-v_i^2, \quad a\,=\,\frac{g^2}{8\pi},
$$
the integral over auxiliary fields yields
\begin{eqnarray}
S[\pi_0]\,&=&\,{\cal N}\,\int \prod_1^2 D\alpha_i\,
e^{\frac 12 i \int_{-\infty}^0
dt_i\,\kappa_i \alpha_i(t)}\,Det F_v^{-\frac 12}\,\times \nonumber \\
&\times&\exp \biggl\{\frac 12 g^2\int dt_1 dt_2 \sum_{i,k} \vec\pi_i^0(t_1)
\,F_{v,ij}^{-1}(t_1,t_2)\,\vec\pi_j^0(t_2)\biggr\}.\nonumber
\end{eqnarray}
The limit $\beta_i\to 0$ (or $r\to 0$) essentially simplifies this answer,
and the $S$-matrix is
$$
S[\pi_0]\,=\,{\cal N}^\prime\,\exp \biggl\{ \frac 12 g^2\,
\sum_{i=1}^2\,\beta_i\,
\int dt_1 dt_2\, \vec \pi_i^0(t_1)\,\bigl[\frac{a}r +K\bigr]^{-1}(t_1,t_2)
\,\vec \pi_j^0(t_2)\biggr\},
$$
or, for small $r$,
$$
S[\pi_0]\,=\,{\cal N}^\prime\,\exp \biggl\{4\pi r\,\sum_{i=1}^2
\beta_i\int_{-\infty}^0 dt\,\vec \pi_i^0(t) \vec \pi_i^0(t)\,\biggr\}.
$$
Here the terms independent of $\pi^0$, in particular the integrals over
$\alpha_i$, are absorbed by the coefficient ${\cal N}^\prime$.
This expression shows that the $S$-matrix turns into a constant at
$r=0$.  Thus, like in the nonrelativistic case, the loop corrections
suppress the soft pion emission. The suppression is resulted both from
the ultraviolet cutoff parameter $r$ and kinematical factors
$\beta_i$.

The above result remains  valid for a more general case.
First, consider
the nonrelativistic $2\to 2$ baryon scattering. Taking the factors
(\ref{fin}) and (\ref{fout}) for the incoming and outgoing baryons,
respectively, the pion
$S$-matrix for this process is
\begin{eqnarray}
\label{S22}
S[\pi_0]\,=\,&{\cal N}&\,\int\prod_{i=1}^2 Dn_i D\alpha_i\,e^{-i
\int_{-\infty}^0 dt_i\bigl[ \frac 12 \alpha_i\kappa_i(\vec n_i^2-1) -
g\vec n_i\vec\pi_i^0 \bigr]} \times  \\
&\times&\,\int\prod_{k=3}^4 Dn_k D\alpha_k\,e^{-i
\int_0^\infty dt_i\bigl[ \frac 12 \alpha_k\kappa_k(\vec n_i^2-1) -
g\vec n_k\vec\pi_k^0 \bigr]}\,e^\Phi,\nonumber
\end{eqnarray}
where
$$
\Phi\,=\, \frac 12 g^2\,\vec J\,\frac 1{\partial^2}\,\vec J,
$$
$$
\vec J\,=\,\theta(-t)\,\sum_{i=1}^2 \vec n_i(t)\,\delta_v^3(x-v_it)\,+\,
\theta(t)\,\sum_{f=3}^4 \vec n_f(t)\,\delta_v^3(x-v_ft).
$$
Denoting $\vec n_i^+(t)=\theta(t)\vec n_i(t)$ and $\vec n_f^-(t)=\theta(-t)
\vec n_f(t)$, the part of an action, which
is quadratic in  $\vec
n(t)$, can be written as $$ \frac 12 \int_0^\infty dt_1 dt_2 \sum_{i,j}
\vec n_i^\sigma(t_1)\, F_{ij}^{\sigma \sigma^\prime}(t_1,t_2)\,
\vec n_j^{\sigma^\prime}(t_2),\quad \sigma, \sigma^\prime = +,-,
$$
where the set of the functions $F_{ij}^{\sigma \sigma^\prime}$ is again
represented by the $2\times 2$
block matrix in the indices $\sigma, \sigma^\prime$:
$$
F\,=\,\left(\begin{array}{cc}F_{11}&F_{12}\\F_{21}&F_{22}\end{array}
\right),
$$
but now each block is itself the $2\times 2$ block matrix in the indices $i,j$.
The blocks $F_{11}$ and $F_{22}$ are given by eq. (\ref{block}) (the
variables $\alpha_{1,2}$ and $\kappa_{1,2}$ in
the block $F_{11}$ for the incoming particles have to be replaced in the
block $F_{22}$ by the variables
$\alpha_{3,4}$ and $\kappa_{3,4}$ for  outgoing ones).

The non-diagonal blocks are
$$
F_{12}\,=\,F_{21}\,=\,\frac{g^2}{8\pi^2}\,\frac 1{(t_1+t_2)^2+i\delta}\,
\left(\begin{array}{cc}I&I\\I&I\end{array}\right),
$$
where the baryon velocities squared are neglected in the
nonrelativistic limit. This expression becomes singular at
$t_1=t_2=0$ only, namely at the central region of the reaction that
is beyond the approximation used here and should be
excluded. Therefore, only the diagonal blocks are singular, when the
regulator $r\to 0$.

The gaussian integral over the fields $n_i^\sigma(t)$  results
again in the
formula (\ref{gauss}), where the sum runs over the indices $i,\sigma$.
Since the matrix $F^{-1}$ can be expressed as
$$
F^{-1}\,=\,\left(\begin{array}{cc}F_{11}^{-1}&0\\0&F_{22}^{-1}\end{array}
\right)\,\left(\begin{array}{cc}I&F_{12}F_{22}^{-1}\\F_{21}F_{11}^{-1}&I
\end{array}\right)^{-1}
$$
and due to the property $F_{12}F_{22}^{-1}=F_{21}F_{11}^{-1}=0$,
which follows
 from  eq. (\ref{Fr}) for  inverse  blocks, the
functional integral over $\alpha$ in eq. (\ref{S22}) decays at $r\to 0$
into the product of the two independent integrals for the incoming and
outgoing baryons.  Both of them result in the constants, so the
$S$-matrix turns out to be constant too.

Since in the ultra-relativistic limit only the diagonal blocks,
$F_{11},F_{22}$, are singular for the small $r$ in the matrix $F$ and they
have the form given by eq. (\ref{Fv}), the integral (\ref{S22}) results
into the $S$-matrix
\begin{equation}
\label{SUR}
S[\pi_0]\,=\,{\cal N}\,\exp \biggl\{4\pi r\,\biggl[\sum_i \beta_i
\int_{-\infty}^0 dt\,\vec \pi_i^0(t) \vec \pi_i^0(t)\,+\,
\sum_f \beta_f
\int_0^\infty dt\,\vec \pi_f^0(t) \vec \pi_f^0(t)\biggr]\,\biggr\},
\end{equation}
which turns  into a constant for $r\to 0$ as well.

\smallskip
\noindent
{\bf 6.} As the two limiting cases lead to the same answer,
it is natural to assume
the pion emission to be suppressed for all kinematics. This conclusion
agrees with the old result \cite{W} and exhibits a general property of loop
corrections to reduce tree amplitudes. The reason for the suppression lies
in the essentially nonlinear character of the pion field which is actually
a phase of the chiral condensate, so the strong field associated with
a large number of pions does not really contribute.

The suppression occurs when the ultraviolet cutoff $r\to 0$, that is when
the loop momenta are formally allowed to go to infinity. However,
an effective
low-energy theory can originally imply some cutoff. From this viewpoint
the expression (\ref{SUR}) can be used for the small finite values of $r$.
Notice that the exponential factors in the
$S$-matrix include the
quadratic combination of  pion fields instead
of the linear one peculiar to a usual coherent state like in
electrodynamics. An opposite limit, $r\to\infty$, kills the loops because of
the absence of infrared singularities for soft pions, and the $S$-matrix
is given by the tree formulae (\ref{Sin}), (\ref{Sout}).

One has to stress the point, at last, that the suppression
occurs only for pions emitted off the "ends", that is from the most
peripheral parts of the process. However, the pion radiation treatment
at the hard stage of the reaction is beyond the present approach, since
it needs a detailed knowledge of the baryon structure at small
distances.

\vspace{0.5cm}
\noindent
{\bf Acknowledgements:} The author is grateful to M.Ryskin
for helpful discussions
and to DESY theory department for hospitality.
\\
This work is financially supported by RFFI grant 96-02-18013.

\end{document}